\documentclass[12pt]{JHEP3}

\usepackage{epsfig,latexsym,amssymb,cite}
\usepackage{amsmath}
\usepackage{graphicx}
\usepackage{bm}

\newcommand{\be}{\begin{equation}}
\newcommand{\ee}{\end{equation}}
\newcommand{\bea}{\begin{eqnarray}}
\newcommand{\eea}{\end{eqnarray}}
\newcommand{\bem}{\begin{multline}}
\newcommand{\eem}{\end{multline}}
\newcommand{\beg}{\begin{gather}}
\newcommand{\eeg}{\end{gather}}

\newcommand{\stackeven}[2]{{{}_{\displaystyle{#1}}\atop\displaystyle{#2}}}
\newcommand{\lsim}{\stackeven{<}{\sim}}

\def\eq#1{{Eq.~(\ref{#1})}}
\def\fig#1{{Fig.~\ref{#1}}}
\newcommand{\ben}{\begin{eqnarray*}}
\newcommand{\een}{\end{eqnarray*}}

\graphicspath{{Figs/}}

\title{Toward Thermalization in Heavy Ion Collisions at Strong
  Coupling}

\author{Yuri V. Kovchegov${}^{\,1}$, Shu Lin${}^{\,2}$ \\
\vspace{0.1in}

${}^{\,1}$Department of Physics, The Ohio State University, Columbus,
OH 43210, USA \vspace{0.1in}

${}^{\,2}$Department of Physics and Astronomy, Stony Brook
  University, Stony Brook NY 11794, USA \\~~\\ 
E-mail addresses: \email{yuri@mps.ohio-state.edu}, 
\email{slin@grad.physics.sunysb.edu} 
\vspace{0.1in}
}

\date{November 2009}

\abstract{We find the trapped surface for a collision of two
  sourceless shock waves in AdS$_5$ and conclude that such collisions
  always lead to a creation of a black hole in the bulk.  Due to
  holographic correspondence, in the boundary gauge theory this result
  proves that a thermalized medium (quark-gluon plasma) is produced in
  heavy ion collisions at strong coupling (albeit in ${\cal N} =4$
  super-Yang-Mills theory). We present new evidence supporting the
  analytic estimate for the time of thermalization that exists in the
  literature and find that thermalization time is parametrically much
  shorter than the time of shock wave stopping, indicating that our
  result may be relevant for description of heavy ion collision
  experiments.}


\preprint{}

\begin{document}



\section{Introduction}

The problem of understanding the physics behind thermalization of the
medium produced in ultrarelativistic heavy ion collisions is one of
the main open questions in heavy ion theory. It has become especially
important in recent years after hydrodynamic simulations indicated
that a very short thermalization time of the order of $1$~fm/c is
required to describe RHIC data \cite{Huovinen:2001cy,Teaney:2001av}.
Lately the problem of thermalization has been studied in the strong
coupling framework of the Anti-de Sitter space/conformal field theory
(AdS/CFT) correspondence
\cite{Maldacena:1997re,Gubser:1998bc,Witten:1998qj} with the goal of
learning about the dynamics of the strongly-coupled QCD medium by
studying the strongly coupled medium in ${\cal N} =4$ super-Yang-Mills
(SYM) theory
\cite{Janik:2005zt,Kajantie:2008rx,Gubser:2008pc,Grumiller:2008va,Albacete:2008vs,Lin:2009pn,Gubser:2009sx,AlvarezGaume:2008fx,Nastase:2008hw}.

The 5-dimensional gravity dual of a shock wave (ultrarelativistic
nucleus) in our 4-dimensional space-time was first constructed in
\cite{Janik:2005zt}. The AdS$_5$ shock wave metrics are shown below in
Eqs. (\ref{nuc1}) and (\ref{nuc2}). These metrics are solutions of
Einstein equations in the AdS$_5$ bulk without sources. In four
dimensions they correspond to nuclei of infinite transverse extent
with a uniform distribution of matter in the transverse plane.
Collisions of shock waves in Eqs.  (\ref{nuc1}) and (\ref{nuc2}) have
been studied in
\cite{Kajantie:2008rx,Grumiller:2008va,Albacete:2008vs,Albacete:2009ji}
with the goal of explicitly constructing the metric after the
collision. While the shock wave collisions in AdS$_3$ allowed for an
exact solution of the problem \cite{Kajantie:2008rx}, it turned out to
be significantly harder in AdS$_5$, allowing only for a perturbative
solution of Einstein equations in graviton exchanges
\cite{Grumiller:2008va,Albacete:2008vs}. While all-order graviton
exchanges with one of the shock waves (corresponding to proton-nucleus
collisions) were resummed exactly in \cite{Albacete:2009ji}, the full
problem of nucleus-nucleus collisions involving all-order graviton
exchanges with {\sl both} nuclei still remains unsolved in AdS$_5$.

In \cite{Gubser:2008pc} an alternative to the exact solution of
Einstein equations was proposed: the authors of \cite{Gubser:2008pc}
constructed a trapped surface for a head-on collision of two shock
waves with sources in the AdS bulk following
\cite{Penrose,Eardley:2002re}.  Sources in the bulk lead to the nuclei
in the boundary theory having some transverse coordinate dependence in
their matter distributions.  Formation of a trapped surface before the
collision indicates that a black hole will be formed in the future,
after the collision. Thus the authors of \cite{Gubser:2008pc} have
proven that black hole is formed in a collision of two shock waves
with point sources in the bulk. Generalizations of
\cite{Gubser:2008pc} to the case of nuclear collisions with non-zero
impact parameter were presented in \cite{Lin:2009pn,Gubser:2009sx}.
Also a trapped surface was found in \cite{Lin:2009pn} for an important
case of collision of two shock waves with extended (not point-like)
bulk sources.

However, the exact implications of a source in the bulk for the
boundary theory are still not entirely clear. The same energy-momentum
tensor of the boundary theory can be given by metrics with extended
sources at different bulk locations. It is possible that the sources
would manifest themselves in fluctuations of the metric, but more
research is needed to understand which bulk source gives the ``right''
fluctuations most accurately describing real-life heavy ion
collisions. In \cite{Lin:2009pn} it was suggested that the position of
the source in the bulk is related to the saturation scale of the shock
wave. Initial steps on determination of saturation scale in shock
waves were done in \cite{Albacete:2008ze,Mueller:2008bt,Avsar:2009xf}.
It appears more work is needed to clarify the complete impact of the
bulk source on the boundary gauge theory.

Interestingly the trapped surfaces found in
\cite{Gubser:2008pc,Lin:2009pn,Gubser:2009sx} are always formed around
the source in the bulk. One may therefore wonder whether the source is
required for the trapped surface to form. No trapped surface analysis
has been performed to date for the sourceless shock waves of Eqs.
(\ref{nuc1}) and (\ref{nuc2}) to answer this question.

Here we perform a trapped surface analysis for a collision of two
sourceless shock waves from Eqs. (\ref{nuc1}) and (\ref{nuc2}). We
first consider the trapped surface obtained in \cite{Lin:2009pn} for a
collision of two shock waves with extended sources in the bulk, and
then take the limit in which the sources are moved to the deep
infrared (IR) while keeping the energies of the shock waves in the
boundary gauge theory fixed.  Interestingly enough, the trapped
surface does not disappear in this source-free limit, its lower
boundary remains at finite value of the 5th dimension coordinate $z$
with its finite area giving a finite expression for the produced
entropy. We argue that collisions of two shock waves with sources in
the deep IR (at $z=\infty$) are indistinguishable from collisions of
two shock waves without bulk sources by performing a perturbative
solution of Einstein equations for the shock wave with sources in the
bulk and taking the sources to $z=\infty$. We also note that the
trapped surface which remains after we send the sources to $z=\infty$
does not depend on how the limit was taken and on which sources were
sent to infinity: the remaining trapped surface is the same for
extended and point-like sources sent to the IR.

We therefore conclude that a collisions of two sourceless shock waves
in AdS$_5$ leads to creation of a black hole in the bulk.  The absence
of bulk sources leaves no uncertainty in the interpretation of the
physics and makes application of AdS/CFT correspondence better
justified. For the boundary theory this result proves that thermalized
quark-gluon plasma is produced in heavy ion collisions at strong
coupling.

The paper is structured as follows. In Sect. \ref{problem} we present
the problem at hand and describe how the limit of sending the sources
to the IR should be taken without changing the bulk physics. In Sect.
\ref{pertsol} we present a lowest-order perturbative solution of
Einstein equations for a collision of two shock waves with sources
along the lines of a similar calculation for the sourceless shock
waves in \cite{Albacete:2008vs}. We take the limit of the shock waves
sources going to the IR and show that our solution exactly maps onto
the metric produced in a collision of two sourceless shock waves found
in \cite{Albacete:2008vs}. This provides a strong argument that the
shock waves with sources at $z=\infty$ collide in the same way as the
shocks without any sources. In Sect. \ref{trap} we perform the trapped
surface analysis and demonstrate that the trapped surface does not
disappear when the sources are send to the deep IR. Thus we obtain the
trapped surface for the collision of two sourceless shock waves. In
Sect. \ref{conc} we conclude by presenting a guess for the
thermalization time inspired by our analysis (see also
\cite{Grumiller:2008va}). We argue that thermalization proper time is
likely to be parametrically shorter than the light-cone stopping time
for shock waves found in \cite{Albacete:2008vs,Albacete:2009ji}, which
indicates that our conclusions may be applied to real-life heavy ion
collisions at least at the qualitative level. We note however that the
numbers generated by our approximate thermalization time estimate are
too short to describe RHIC physics.


\section{The Problem}
\label{problem}

High energy heavy ion collision can be realistically modeled by a
collision of two ultrarelativistic shock waves.  In
\cite{Janik:2005zt}, using the holographic correspondence
\cite{deHaro:2000xn}, the geometry in AdS$_5$ dual to each one of the
nuclei in the boundary theory is given by the following metric
\begin{align}\label{nuc1}
  ds^2 \, = \, \frac{L^2}{z^2} \, \left\{ -2 \, dx^+ \, dx^- + t_1
    (x^-) \, z^4 \, d x^{- \, 2} + d x_\perp^2 + d z^2 \right\}
\end{align}
for nucleus 1 and by 
\begin{align}\label{nuc2}
  ds^2 \, = \, \frac{L^2}{z^2} \, \left\{ -2 \, dx^+ \, dx^- + t_2
    (x^+) \, z^4 \, d x^{+ \, 2} + d x_\perp^2 + d z^2 \right\}
\end{align}
for nucleus 2. Here $d x_\perp^2 = (d x^1 )^2 + (d x^2)^2$ is the
transverse metric and $x^\pm = (x^0 \pm x^3) / \sqrt{2}$ where $x^3$
is the collision axis. $L$ is the radius of S$_5$ and $z$ is the
coordinate describing the 5th dimension with the boundary of AdS$_5$
at $z=0$. We have also defined
\begin{align}\label{tt}
  t_1 (x^-) \, \equiv \, \frac{2 \, \pi^2}{N_c^2} \, \langle T_{1 \,
    --} (x^-) \rangle, \ t_2 (x^+) \, \equiv \, \frac{2 \,
    \pi^2}{N_c^2} \, \langle T_{2 \, ++} (x^+) \rangle
\end{align}
in accordance with the prescription of holographic renormalization
\cite{deHaro:2000xn}. Here $\langle T_{1 \, --} (x^-) \rangle$ and
$\langle T_{2 \, ++} (x^+) \rangle$ are the energy-momentum tensors of
the two shock waves in the gauge theory. We assume that the nuclei are
so large and homogeneous that one can neglect transverse coordinate
dependence in $\langle T_{1 \, --} (x^-) \rangle$ and $\langle T_{2 \,
  ++} (x^+) \rangle$. Following \cite{Janik:2005zt} we take
\begin{align}\label{emtshock}
  \langle T_{1 \, --} (x^-) \rangle \, = \, \mu_1 \, \delta (x^-), \ 
  \langle T_{2 \, ++} (x^+) \rangle \, = \, \mu_2 \, \delta (x^+).
\end{align}
For simplicity we also put $\mu_1 = \mu_2 = \mu$. 

The metrics in Eqs. (\ref{nuc1}) and (\ref{nuc2}) solve Einstein
equations in the empty AdS$_5$ space:
\begin{align}\label{Ein0}
  R_{\mu\nu} - \frac{1}{2} \, g_{\mu\nu} \, R - \frac{6}{L^2} \,
  g_{\mu\nu} \, = \, 0.
\end{align}
However, as we will see below, it is hard to perform the trapped
surface analysis with the sourceless shock waves. To this end, as we
have mentioned above, it will be more convenient to represent
sourceless shock waves as limiting cases of the shock waves with
sources, when the sources are sent to $z=\infty$ while keeping
energy-momentum tensor of the nuclei in the boundary theory intact.

We therefore need to construct shock waves with sources in the bulk,
which we will do following \cite{Lin:2009pn,Gubser:2008pc}. We need to
satisfy Einstein equations in AdS$_5$ with sources in the bulk
\begin{align}\label{Ein1}
   R_{\mu\nu} - \frac{1}{2} \, g_{\mu\nu} \, R - \frac{6}{L^2} \,
  g_{\mu\nu} \, = \, 8 \, \pi \, G_5 \, J_{\mu\nu}
\end{align}
where $J_{\mu\nu}$ is the energy momentum tensor for bulk sources. We
will not specify what fields contribute to create non-zero
$J_{\mu\nu}$ in the bulk: as for us the source will serve as an IR
regulator we do not need to know the origin of $J_{\mu\nu}$ in detail.
The 5-dimensional Newton constant is
\begin{align}
  G_5 = \frac{\pi \, L^3}{2 \, N_c^2}.
\end{align}
\eq{Ein1} can be rewritten as
\begin{align}\label{Ein2}
  R_{\mu\nu} + \frac{4}{L^2} \, g_{\mu\nu} \, = \, 8 \, \pi \, G_5 \,
  \left( J_{\mu\nu} - \frac{1}{3} \, g_{\mu\nu} \, J \right)
\end{align}
with 
\begin{align}
  J \, = \, J_\mu^{\ \mu} \, = \, J_{\mu\nu} \, g^{\mu\nu}.
\end{align}

Following \cite{Lin:2009pn} for one shock wave we will consider a
source without any transverse ($x^1, x^2$) coordinate dependence, with
the only non-zero component of the energy-momentum tensor
\begin{align}\label{source0}
  J^{(0)}_{--} \, = \, \frac{E}{z_0 \, L} \, \delta (x^-) \, \delta (z
  - z_0).
\end{align}
The source is located at $z = z_0$ and $x^- =0$, spans the transverse
directions and moves along the $x^+$ axis. $E$ is a yet unspecified
parameter with dimension of energy.  To find the metric of the shock
wave satisfying \eq{Ein2} with the source (\ref{source0}) we look for
it in the following form generalizing \eq{nuc1}
\begin{align}\label{nuc1s}
  ds^2 \, = \, \frac{L^2}{z^2} \, \left\{ -2 \, dx^+ \, dx^- + \phi
    (z) \, \delta (x^-) \, d x^{- \, 2} + d x_\perp^2 + d z^2
  \right\}.
\end{align}
Plugging Eqs. (\ref{nuc1s}) and (\ref{source0}) into \eq{Ein2} we get
the following equation for the ``$--$'' component of Einstein
equations \cite{Lin:2009pn}
\begin{align}\label{phi_eq}
  \frac{3}{2 \, z} \, \phi' (z) - \frac{1}{2} \, \phi'' (z) \, = \, 8
  \, \pi \, G_5 \, \frac{E}{z_0 \, L} \, \delta (z - z_0).
\end{align}
When solving this equation we require that $\phi (z) \rightarrow 0$ as
$z \rightarrow 0$ and that $\phi (z)$ is regular as $z \rightarrow +
\infty$. The latter condition is needed to avoid the singular behavior
of metrics (\ref{nuc1}) and (\ref{nuc2}) in the IR. While the
singularity of metrics (\ref{nuc1}) and (\ref{nuc2}) does not 
affect curvature invariants and is thus not unphysical, it is easier to
perform trapped surface analysis which we intend to do below on a
metric with is regular in the IR.

Solving \eq{phi_eq} with the boundary condition that $\phi (z)
\rightarrow 0$ as $z \rightarrow 0$ and $\phi (z)$ is regular at $z
\rightarrow + \infty$ yields \cite{Lin:2009pn}
\begin{align}\label{phi_sol}
  \phi (z) \, = \, \frac{4 \, \pi \, G_5 \, E}{L} \, \left\{
    \begin{array}{c}
      \frac{z^4}{z_0^4}, \ z \le z_0 \\~\\
        1, \ z > z_0.
    \end{array}
\right.
\end{align}
Eqs. (\ref{phi_sol}) and (\ref{nuc1s}) give us the metric of a single
shock wave with the bulk source (\ref{source0}). 

Using holographic renormalization \cite{deHaro:2000xn} (see e.g.
\eq{tt} above) we conclude that the energy-momentum tensor
corresponding to the metric (\ref{nuc1s}) has only one non-zero
component \cite{Lin:2009pn}
\begin{align}
  \langle T_{--} \rangle \, = \, \frac{L^3}{4 \, \pi \, G_5} \, \delta
  (x^-) \, \lim_{z \rightarrow 0} \frac{\phi (z)}{z^4} \, = \, \frac{E
    \, L^2}{z_0^4} \, \delta (x^-).
\end{align}
It is clear that this energy-momentum tensor would be the same as for
the sourceless shock wave (\ref{nuc1}) given by \eq{emtshock} if we
identify
\begin{align}\label{mu}
  \mu \, = \, \frac{E \, L^2}{z_0^4}
\end{align}
obtaining
\begin{align}\label{emt1}
  \langle T_{--} \rangle \, = \, \mu \, \delta (x^-).
\end{align}

The difference between the metrics for the shock wave with source in
Eqs. (\ref{phi_sol}) and (\ref{nuc1s}) and the sourceless shock wave
in \eq{nuc1} is that the source regulates the metric in the IR. It is
important to note that if we take $z_0 \rightarrow \infty$ limit of
the metric in Eqs. (\ref{nuc1s}) and (\ref{phi_sol}) keeping $E/z_0^4$
(and therefore $\mu$ in \eq{mu} fixed) we would recover the metric in
\eq{nuc1} without modifying the energy-momentum tensor of the gauge
theory given by (\ref{emt1}). At any finite $z$ the metric of Eqs.
(\ref{nuc1s}) and (\ref{phi_sol}) becomes equivalent to (\ref{nuc1})
in this limit, which sends the source at $z_0$ to the IR infinity. The
question arises whether the metric (\ref{nuc1}) is equivalent to the
$z_0 \rightarrow \infty$, $E/z_0^4 = const$ limit of the metric in
Eqs.  (\ref{nuc1s}) and (\ref{phi_sol}). In other words, is having the
sources at infinity identical to having no sources at all?

We are interested in the answer to this question in the context of
collisions of two shock waves. The question then becomes whether
colliding shock waves from Eqs. (\ref{nuc1}) and (\ref{nuc2}) are
identical to colliding the shock wave in Eqs.  (\ref{nuc1s}) and
(\ref{phi_sol}) with its counterpart with $x^+ \leftrightarrow x^-$
in the limit $z_0 \rightarrow \infty$, $E/z_0^4 = const$ of
the resulting post-collision metric?

The intuitive answer to the this question is ``yes''. Indeed it is
highly unlikely that sources at $z_0 = \infty$ would affect any
physics at finite $z$. Even in empty AdS$_5$ space light propagates
with velocity 1 along the $z$-direction. It would take light an
infinite time to travel to any finite $z$ from $z_0 = \infty$ after
the collision. The metric modification in the collision is only likely
to lower the light velocity in the $z$-direction: in the ``extreme''
case when a black hole is created no signal from $z = \infty$ would be
able to propagate outside of the horizon. Even more minor
modifications of the metric are likely to only change the speed of
light in $z$-direction leaving it finite and not changing the above
arguments. Hence any modification of sources at $z_0 = \infty$ in the
collision is not going to affect the physics at finite $z$. Hence the
collision of two shock waves with sources at $z_0 = \infty$ should be
indistinguishable from the collision of two sourceless shock waves in
Eqs. (\ref{nuc1}) and (\ref{nuc2}).

One may also think of a source at $z_0$ as providing an (externally
imposed) infrared cutoff $1/z_0$ on the transverse momenta $k_T$ of
the partons inside the shock wave in the boundary gauge theory (see
\cite{Lin:2009pn}). With this interpretation the limit of $z_0
\rightarrow \infty$, $E/z_0^4 = const$ can be interpreted in the
boundary theory as removing the IR cutoff on the transverse momenta of
the partons while keeping the energy of the shock wave fixed. The
shock waves without sources would then correspond to nuclei without an
ad hoc IR cutoff on the transverse momenta of their partons in the
boundary theory. Hence, from the standpoint of the boundary theory,
the $z_0 \rightarrow \infty$ limit imposed on the four-dimensional
shock waves dual to the shocks with sources in the bulk would simply
remove the IR cutoff on partons' $k_T$. This would make the boundary
theory shock waves identical to those dual to the sourceless shock
waves in the bulk. Therefore, with the IR $k_T$-cutoff interpretation
of $1/z_0$ \cite{Lin:2009pn} the $z_0 \rightarrow \infty$ limit also
appears to be a justified way of obtaining duals of sourceless bulk
shock waves in the boundary theory.

To verify the above arguments we will perform a perturbative solution
of Einstein equations for a collision of two shock waves with sources
in the next Section. We will explicitly demonstrate that taking the
$z_0 \rightarrow \infty$, $E/z_0^4 = const$ limit of the obtained
metric produced in the collision would simply reduce it to the metric
produced in the collision of two sourceless shock waves found
previously in \cite{Albacete:2008vs,Grumiller:2008va}, thus
substantiating our intuitive argument above.


\section{Perturbative Solution of Einstein Equations for Colliding 
  Shock \\ Waves with Bulk Sources}

\label{pertsol}

Consider a collision of two shock waves with sources like the one
given in \eq{source0}. The general metric for such a collision could
be written as
\begin{align}\label{AA_gen}
  ds^2 \, = \, \frac{L^2}{z^2} \, \bigg\{ -\left[ 2 + g (x^+, x^-, z)
  \right] \, dx^+ \, dx^- + \left[ \phi(z) \, \delta (x^-) + f (x^+,
    x^-, z) \right] \, d x^{- \, 2} \notag \\ + \left[ \phi (z) \,
    \delta (x^+) + {\tilde f} (x^+, x^-, z) \right] \, d x^{+ \, 2} +
  \left[ 1 + h (x^+, x^-, z) \right] \, d x_\perp^2 + d z^2 \bigg\}.
\end{align}
The functions $f$, $\tilde f$, $g$, and $h$ are non-zero only for $x^+
\ge 0$, $x^- \ge 0$. Before the collision (for $x^- < 0$ and $x^+ <0$)
the superposition of the metrics of colliding shocks (the terms with
$\phi$'s above) solves Einstein equations (\ref{Ein2}) exactly. 

We will follow \cite{Albacete:2008vs,Grumiller:2008va,Albacete:2009ji}
and find the functions $f$, $\tilde f$, $g$, and $h$ perturbatively at
the lowest order treating the shock waves as perturbations of the
empty AdS$_5$ space. As $\phi (z) \sim \mu$ one can argue that $f$,
$\tilde f$, $g$, and $h$ start at order $\mu^2$
\cite{Albacete:2008vs,Albacete:2009ji}.  Our strategy is to expand
Einstein equations to the order linear in $f$, $\tilde f$, $g$, and
$h$ and quadratic in $\phi$. This is the same procedure as used in
\cite{Albacete:2008vs,Albacete:2009ji} for a collision of two
sourceless shock waves. 

The main difference in the case at hand is that the shock waves now
have sources. The energy-momentum tensors of the sources, given by the
following non-vanishing components before the collision (order $\mu$,
see Eqs. (\ref{mu} and (\ref{source0})))
\begin{align}\label{source02}
  J^{(0)}_{--} \, = \, \mu \, \frac{z_0^3}{L^3} \, \delta (x^-) \,
  \delta (z - z_0), \ \ \ J^{(0)}_{++} \, = \, \mu \,
  \frac{z_0^3}{L^3} \, \delta (x^+) \, \delta (z - z_0),
\end{align}
get modified in the collision. In principle to understand
modifications of the bulk source one needs to know the field content of
the source and the corresponding equations of motion for the fields.
However, it turns out that this is not really necessary. Following a
similar procedure for perturbative construction of classical
Yang-Mills fields in nuclear collisions \cite{Kovchegov:1997ke} we
note that Einstein equations (\ref{Ein1}) imply
\begin{align}\label{emtcon}
  \nabla_\mu \, J^{\mu\nu} \, = \, 0
\end{align}
where $\nabla_\mu$ is the covariant derivative. Imposing causality and
using \eq{emtcon} along with Einstein equations one can perturbatively
construct the bulk energy-momentum tensor order-by-order in $\mu$.
Using the symmetries of the problem one can argue that it is unlikely
that colliding sources would recoil in the transverse or $z$
directions. This limits the non-zero contributions to the bulk
energy-momentum tensor to $J_{++}$, $J_{--}$ and $J_{+-} \, = \,
J_{-+}$.  Note that to find $J^{\mu\nu}$ at order $\mu^2$ one only
need the metric (\ref{AA_gen}) at order $\mu$. This means one does not
yet need to know the functions $f$, $\tilde f$, $g$, and $h$. It is
then not too hard to infer the sources up to order $\mu^2$: the
non-vanishing components of the bulk energy-momentum tensor are
\begin{subequations}\label{source}
\begin{align}
  J_{++} \, & = \, \mu \, \frac{z_0^3}{L^3} \, \delta (z - z_0) \,
  \left[ \delta (x^+) + \, \frac{1}{2} \, \theta
    (x^-) \, \delta' (x^+) \, [ z \, \phi' (z) - \phi (z) ] + \ldots \right] \\
  J_{--} \, & = \, \mu \, \frac{z_0^3}{L^3} \, \delta (z - z_0) \,
  \left[ \delta (x^-) \, + \, \frac{1}{2} \, \theta
    (x^+) \, \delta' (x^-) \, [z \, \phi' (z) - \phi (z) ] + \ldots \right] \\
  J_{+-} \, & = \, J_{-+} \, = \, - \mu \, \frac{z_0^3}{L^3} \, \delta
  (z - z_0) \, \delta (x^+) \, \delta (x^-) \, \left[ \phi (z) +
    \frac{1}{2} \, z \, \phi' (z) \right] + \ldots \, .
\end{align}
\end{subequations}

Plugging Eqs. (\ref{AA_gen}) and (\ref{source}) into (\ref{Ein2}) and
expanding the result in powers of $\mu$ we obtain at order $\mu^2$ the
following expressions for the ``$\bot \bot$'' and the ``$zz$''
components of Einstein equations
\begin{subequations}\label{ein_AA}
\begin{align}
  (\bot \bot) \hspace*{0.15in} g_z + 5 \, h_z - z \, h_{z\, z} + 2 \,
  z \, h_{x^+ \, x^-} \, = & \ 2 \, \delta (x^+) \, \delta (x^-) \,
  \phi (z) \, \phi' (z) \notag \\ & - \frac{16 \, \pi}{3} \,
  \frac{G_5}{L^3} \, z_0^5 \, \mu \, \delta (x^+) \, \delta (x^-) \,
  \delta
  (z - z_0) \, \phi' (z) \label{pp} \\
  (zz) \hspace*{0.15in} g_z + 2 \, h_z - z \, g_{z\, z} - 2 \, z \,
  h_{z\, z} = & - \delta (x^+) \, \delta (x^-) \, \left[ - 2 \, \phi
    (z) \, \phi' (z) + z \, ( \phi' (z) )^2 + 2 \, z \, \phi (z) \,
    \phi'' (z) \right] \notag \\ & - \frac{16 \, \pi}{3} \,
  \frac{G_5}{L^3} \, z_0^5 \, \mu \, \delta (x^+) \, \delta (x^-) \,
  \delta (z - z_0) \, \phi' (z). \label{zz}
\end{align}
\end{subequations}
Here the subscripts indicate partial derivatives. Solving \eq{pp} for
$g_z$ and substituting the result into \eq{zz} yields
\begin{align}\label{heq1}
  - 3 \, h_z + 3 \, z \, h_{z \, z} & - z^2 \, h_{z \, z \, z} + 2 \,
  z^2 \, h_{x^+ \, x^- \, z} \, = \, \delta (x^+) \, \delta (x^-)
  \notag \\ \times \, & \left[ z \, [ \phi' (z) ]^2 - \frac{16 \,
      \pi}{3} \, \frac{G_5}{L^3} \, z_0^5 \, \mu \, z \, \left[
      \delta' (z - z_0) \, \phi' (z) + \delta (z - z_0) \, \phi'' (z)
    \right] \right]. 
\end{align}
\eq{heq1} can be rewritten as
\begin{align}\label{heq2}
  z^2 \, \partial_z \, & \left[ \frac{3}{z} \, h_z - h_{z \, z} + 2 \,
    h_{x^+ \, x^-} \right] \, = \, \delta (x^+) \, \delta (x^-) \notag
  \\ \times \, & \left[ z \, [ \phi' (z) ]^2 - \frac{16 \, \pi}{3} \,
    \frac{G_5}{L^3} \, z_0^5 \, \mu \, z \, \left[ \delta' (z - z_0)
      \, \phi' (z) + \delta (z - z_0) \, \phi'' (z) \right] \right].
\end{align}
We can now substitute $\phi (z)$ from \eq{phi_sol} into \eq{heq2}.
There is a small subtlety: the derivative of $\phi (z)$ is
discontinuous at $z=z_0$. It is therefore not clear which value of the
derivative to choose, the one at $z - z_0 \rightarrow 0^+$ or the one
at $z - z_0 \rightarrow 0^-$. As for $z>z_0$ all derivatives of $\phi
(z)$ are zero, plugging the derivatives at $z - z_0 \rightarrow 0^+$
into \eq{heq2} would simply eliminate all bulk source effects. It
therefore seems more physical to use the derivatives at $z - z_0
\rightarrow 0^-$. This gives
\begin{align}\label{heq3}
  \frac{3}{z} \, h_z - h_{z \, z} + 2 \, h_{x^+ \, x^-} \, = \,
  \frac{1}{3} \, \left( \frac{16 \, \pi \, G_5 \, \mu}{L^3} \right)^2
  \, \delta (x^+) \, \delta (x^-) \, \left[ \frac{z^6}{2} \, \theta
    (z_0 -z) - \frac{z_0^6}{2} \, \theta (z -z_0) - z_0^7 \, \delta
    (z-z_0) \right].
\end{align}

\eq{heq3} is easy to solve as the Green function for the operator on
its left hand side was found in
\cite{Danielsson:1998wt,Albacete:2009ji}. Defining the Green function
by
\begin{align}\label{G1}
  \left[ \frac{3}{z} \, \partial_z - \partial_z^2 + 2 \, \partial_+ \,
    \partial_- \right] \, G (x^+, x^-, z; x'^+, x'^-, z') \, = \,
  \delta (x^+ - x'^+) \, \delta (x^- - x'^-) \, \delta (z-z')
\end{align}
one can find an integral expression \cite{Danielsson:1998wt,Albacete:2009ji}
\begin{align}\label{G2}
  G (x^+, x^-, z; x'^+, x'^-, z') \, & = \, \frac{1}{2} \, \theta (x^+
  - x'^+) \, \theta (x^- - x'^-) \, \frac{z^2}{z'} \,
  \int\limits_0^\infty d m \notag \\ & \times \, m \, J_0\left( m \,
    \sqrt{2 \, (x^+ - x'^+) \, (x^- - x'^-)} \right) \, J_2 (m \, z)
  \, J_2 (m \, z')
\end{align}
which can be integrated to give
\begin{align}\label{G3}
  G (x^+, x^-, z; x'^+, x'^-, z') \, & = \, \frac{1}{2 \, \pi} \,
  \theta (x^+ - x'^+) \, \theta (x^- - x'^-) \, \theta (s) \, \theta
  (2-s) \, \frac{z}{z'^2} \, \frac{1 + 2 \, s \, (s-2)}{\sqrt{s \,
      (2-s)}}
\end{align}
with
\begin{align}
  s \equiv \frac{2 \, (x^+ - x'^+) \, (x^- - x'^-) - (z-z')^2}{2 \, z
    \, z'}.
\end{align}

With the help of \eq{G2} we solve \eq{heq3} and write
\begin{align}\label{hsol1}
  h (x^+, x^-, z) & = \int\limits_{-\infty}^{x^+} d x'^+
  \int\limits_{-\infty}^{x^-} d x'^- \int\limits_0^\infty d z'
  \frac{z^2}{2 \, z'} \int\limits_0^\infty d m \, m \, J_0\left( m
    \sqrt{2 (x^+ - x'^+) (x^- - x'^-)} \right)
  J_2 (m \, z) J_2 (m \, z')  \notag \\
  & \times \, \frac{1}{3} \, \left( \frac{16 \, \pi \, G_5 \,
      \mu}{L^3} \right)^2 \, \delta (x'^+) \, \delta (x'^-) \, \left[
    \frac{z'^6}{2} \, \theta (z_0 -z') - \frac{z_0^6}{2} \, \theta (z'
    -z_0) - z_0^7 \, \delta (z'-z_0) \right]. 
\end{align}
Integrating over $x'^+$ and $x'^-$ trivially yields
\begin{align}\label{hsol2}
  h (x^+, x^-, z) = \frac{1}{3} \, \left( \frac{16 \, \pi \, G_5 \,
      \mu}{L^3} \right)^2 \, & \theta (x^+) \, \theta (x^-) \,
  \int\limits_0^\infty d z' \frac{z^2}{2 \, z'} \int\limits_0^\infty d
  m \, m \, J_0\left( m \, \tau \right)
  J_2 (m \, z) J_2 (m \, z')  \notag \\
  & \times \, \left[ \frac{z'^6}{2} \, \theta (z_0 -z') -
    \frac{z_0^6}{2} \, \theta (z' -z_0) - z_0^7 \, \delta (z'-z_0)
  \right]
\end{align}
where we defined the proper time
\begin{align}
  \tau \, = \, \sqrt{2 \, x^+ \, x^-}.
\end{align}

Let us evaluate the three terms in the brackets in \eq{hsol2}
separately. Start with the last term: it is proportional to
\begin{align}\label{term3}
  \int\limits_0^\infty d z' \frac{z^2}{2 \, z'} \int\limits_0^\infty d
  & m \, m \, J_0\left( m \, \tau \right) J_2 (m \, z) J_2 (m \, z')
  \, z_0^7 \, \delta (z'-z_0) \, = \, \frac{z^2}{2} \, z_0^6 \,
  \int\limits_0^\infty d m \, m \, J_0\left( m \, \tau \right) J_2 (m
  \, z) J_2 (m \, z_0) \notag \\
  & = \, \frac{z^2}{2 \, \pi} \, z_0^6 \, \theta (s_0) \, \theta
  (2-s_0) \, \frac{1 + 2 \, s_0 \, (s_0-2)}{\sqrt{s_0 \, (2-s_0)}}
\end{align}
with
\begin{align}\label{s0}
  s_0 = \frac{\tau^2 - (z-z_0)^2}{2 \, z \, z_0}.
\end{align}
We see that taking $z_0 \rightarrow \infty$ and keeping $\mu$ fixed
gives $s_0 \approx - z_0 / (2 \, z)$ such that the expression in
\eq{term3} becomes zero due to $\theta (s_0)$. Hence the last term in
the brackets of \eq{hsol2} does not contribute in the $z_0 \rightarrow
\infty$ limit.

The second term in the brackets of \eq{hsol2} is proportional to
\begin{align}
  \int\limits_{z_0}^\infty d z' \frac{1}{z'} \int\limits_0^\infty d m
  \, m \, J_0\left( m \, \tau \right) J_2 (m \, z) J_2 (m \, z') \, =
  \, \frac{1}{z_0} \, \int\limits_0^\infty d m \, J_0\left( m \, \tau
  \right) J_2 (m \, z) J_1 (m \, z_0) \, = \, 0
\end{align}
with the last step being valid for $z_0 > z + \tau$, i.e., for the
large $z_0$ we are interested in.

We are left with the first term in the brackets of \eq{hsol2}. Hence
at large $z_0$ we have
\begin{align}\label{hsol3}
  h (x^+, x^-, z) & = \frac{1}{3} \, \left( \frac{8 \, \pi \, G_5 \,
      \mu}{L^3} \right)^2 \, \theta (x^+) \, \theta (x^-) \, z^2
  \int\limits_0^{z_0} d z' z'^5 \int\limits_0^\infty d m \, m \,
  J_0\left( m \, \tau \right) J_2 (m \, z) J_2 (m \, z') \notag \\ & =
  \frac{1}{3} \, \left( \frac{8 \, \pi \, G_5 \, \mu}{L^3} \right)^2
  \, \theta (x^+) \, \theta (x^-) \, z^2 \, z_0^4 \,
  \int\limits_0^\infty \frac{d m}{m} \, J_0\left( m \, \tau \right)
  J_2 (m \, z) \, \left[ 6 \, J_4 (m z_0) - m z_0 \, J_5 (m z_0)
  \right] \notag \\ & = \left( \frac{8 \, \pi \, G_5 \, \mu}{L^3}
  \right)^2 \, \theta (x^+) \, \theta (x^-) \, z^4 \, \left[ \tau^2 +
    \frac{1}{3} \, z^2 \right].
\end{align}
This is exactly the solution found for sourceless shock waves in
\cite{Albacete:2008vs}! Using $h$ from \eq{hsol3} in \eq{pp} one would
obtain function $g$, which, for $z_0 \rightarrow \infty$ would also be
$z_0$-independent and would also correspond to that found for
sourceless shock waves in \cite{Albacete:2008vs}. Similarly one can
show that $f$ and $\tilde f$ would also reduce to the ones from
\cite{Albacete:2008vs} in the $z_0 \rightarrow \infty$ limit. We
conclude that, at least at this lowest non-trivial order in $\mu$,
colliding shock waves with sources gives a metric which in the limit of
$z_0 \rightarrow \infty$ (keeping $\mu$ fixed) reduces to that
produced in the collision of two shock waves without sources. This
presents a strong argument supporting our earlier assertion that
collisions of the shock waves with sources at $z_0 = \infty$ are
equivalent to collisions of the shock waves without the sources.


\section{Trapped Surface Analysis}
\label{trap}

Below we will present trapped surface analysis for a collision of two
shock waves without bulk sources. We will begin by outlining general
concepts of the trapped surface analysis and will present a naive
attempt to find the trapped surface for a collision of shock waves
from Eqs. (\ref{nuc1}) and (\ref{nuc2}). We will then obtain the
trapped surface for a collision of two shock waves with bulk sources
and take the limit of $z_0 \rightarrow \infty$, deriving the trapped
surface for a collision of sourceless shock waves. We will solidify
our above conclusion of the equivalence between the sourceless shock
wave and the one with sources at $z=\infty$ by taking the limit of
sources going to the IR for a collision of two different shock waves
with extended sources at $z_1$ and $z_2$ and showing that the limiting
trapped surface is the same as obtained before.


\subsection{Generalities}

Let us start with outlining some generalities of trapped surface.
Consider the collision of two shock waves given by the following
metric before the collision:
\begin{align}\label{metrics}
  ds^2 \, = \, \frac{L^2}{z^2}\left\{-2 \,
    dx^+dx^-\,+\,dx_\perp^2\,+\,dz^2\right\}\,+\, \frac{L}{z} \,
  \Phi_1(x_\perp,z) \, \delta(x^+) \, dx^{+2}\,+\,\frac{L}{z} \,
  \Phi_2(x_\perp,z) \, \delta(x^-) \, dx^{-2}
\end{align}
where (cf. \eq{AA_gen})
\begin{align}
\Phi_i(x_\perp,z) \, = \, \frac{L}{z} \, \phi_i(x_\perp,z), \quad i=1,2.
\end{align}

The marginally trapped surface is found from the condition of
vanishing of expansion $\theta$\cite{Hawking:1969sw}. The trapped
surface is made up of two pieces: ${\cal S}={\cal S}_1\cup{\cal S}_2$.
${\cal S}_1({\cal S}_2)$ is associated with shock wave at $x^+=0 \ 
(x^-=0)$ before the collision.  An additional condition is imposed
requiring that the outer null normal to ${\cal S}_1$ and ${\cal S}_2$
must be continuous at the intersection ${\cal C}={\cal S}_1\cap{\cal
  S}_2$ point $x^+=x^-=0$ to avoid delta function in the expansion.

To calculate the trapped surface associated with shock wave at
$x^+=0$, we use the following coordinate transformation
\cite{Gubser:2008pc,Eardley:2002re}:
\begin{align}\label{coordtr}
  x^-\rightarrow x^-+\frac{\phi_1(x_\perp,z)}{2} \, \theta(x^+)
\end{align}
to eliminate the delta-function discontinuity at $x^+
=0$.\footnote{Note a different definition for the light-cone
  coordinates used in \cite{Gubser:2008pc,Eardley:2002re}.}  The
trapped surface ${\cal S}_1$ can then be parametrized by
\cite{Eardley:2002re}
\begin{align}\label{param}
  x^+=0, \ x^-=-\frac{\psi_1(x_\perp,z)}{2}.
\end{align}

The condition of marginally trapped surface is the vanishing of
expansion $\theta\equiv \, h^{\mu\nu}\nabla_{\mu} \, l_{\nu}$, with
$h^{\mu\nu}$ the induced metric and $l_{\nu}$ the outer null normal to
the trapped surface. Similarly to
\cite{Gubser:2008pc,Lin:2009pn,Gubser:2009sx}, the condition gives
rise to
\begin{align}\label{eom1}
  \left(\Box\,-\,\frac{3}{L^2}\right) \, \left[
    \Psi_1(x_\perp,z)-\Phi_1(x_\perp,z) \right] \, = \, 0
\end{align}
with $\Psi_1(x_\perp,z)=\frac{L}{z}\psi_1(x_\perp,z)$ and the Laplacian is defined
with respect to Euclidean $AdS_3$ space
\begin{align}\label{ads3}
  ds^2 \, = \, \frac{L^2}{z^2} \, \left\{dx_\perp^2+dz^2\right\}.
\end{align}
By analogy, we have the condition defining the trapped surface ${\cal S}_2$:
\begin{align}\label{eom2}
  \left(\Box\,- \, \frac{3}{L^2}\right) \, \left[
    \Psi_2(x_\perp,z)-\Phi_2(x_\perp,z) \right] \, = \, 0.
\end{align}

The continuity of trapped surface ${\cal S}_1$ and ${\cal S}_2$ and
their outer null normal on the cusp of the light-cone $x^+=x^-=0$
reduce to the boundary conditions
\begin{subequations}\label{bc}
\begin{align}
  \Psi_1(x_\perp,z)|_{\cal C} \, = \, \Psi_2(x_\perp,z)|_{\cal C} \, = \, 0 \\
  \nabla\Psi_1(x_\perp,z) \cdot \nabla \Psi_2(x_\perp,z)|_{\cal C} \,
  = \, 8
\end{align}
\end{subequations}
where the boundary ${\cal C}$ is to be determined from \eq{bc}. The
covariant derivative $\nabla$ is again defined with respect to
\eq{ads3}.

Having the equations for the trapped surface with arbitrary shock wave
(\ref{eom1}), (\ref{eom2}) and (\ref{bc}) at hand, we are ready to
apply them to the collision of source-free shock waves (\ref{nuc1})
and (\ref{nuc2}).  With the symmetry
$\phi_1(z)=\phi_2(z)\equiv\phi(z)$ (and thus
$\psi_1(z)=\psi_2(z)\equiv\psi(z)$), they take a particularly simple
form
\begin{subequations}\label{free}
\begin{align}
  &z^2 \, \Psi''(z)-z \, \Psi'(z)-3 \, \Psi(z) \, = \, 0\\
  &\Psi(z_a) \, = \, \Psi(z_b) \, = \, 0\label{cs}\\
  &\frac{z_a^2}{L^2} \, \Psi'(z_a)^2 \, = \, \frac{z_b^2}{L^2} \,
  \Psi'(z_b)^2 \, = \, 8.
  \label{cstr}
\end{align}
\end{subequations}
The boundary ${\cal C}$ in this case is given by $z_a<z<z_b$, as there
is no dependence on transverse coordinates. Eq.~(\ref{free}) is easily
solved by
\begin{align}
\Psi (z) \, = \, C_1 \, z^3 + \frac{C_2}{z}
\end{align}
with $C_1$ and $C_2$ arbitrary constants. 

Obviously we cannot have $C_1=C_2=0$ because of \eq{cstr}. It is easy
to see then \eq{cs} would immediately require $z_a=z_b$.  Similar
phenomenon of no trapped surface was observed in \cite{Eardley:2002re}
for collisions of gravitational shock waves in asymptotically
Minkowskian 4-dimensional space-time.  One may be tempted to conclude
that trapped surface formation is not possible in collisions of
source-free shock waves. Before accepting such conclusion, let us
point out that the reason we choose ${\cal C}$ to be bounded by
$z_a<z<z_b$ from both sides in the bulk is because the trapped surface
has to be closed. However, AdS$_5$ is different from asymptotically
Minkowskian spaces: it appears not quite clear whether the requirement
of a closed trapped surface necessarily implies finite
$z_b$.\footnote{Requirement that the trapped surface has to be closed
  appears to stem from the cosmic censorship conjecture, which we
  assume to be true in AdS$_5 \times S_5$: the issue of whether
  trapped surfaces in AdS$_5$ necessarily have to be closed may
  require further investigation.} If one searches for the trapped
surface with $z_b = \infty$, i.e., with $z>z_a$ constraint only, such
that conditions in Eqs.~(\ref{free}) are imposed only at $z_a$, one
gets
\begin{align}\label{Psia}
  \Psi (z) \, = \, \frac{L}{\sqrt{2}} \, \left[ \frac{z^3}{z_a^3} -
    \frac{z_a}{z} \right]
\end{align}
giving
\begin{align}\label{psia}
  \psi (z) \, = \, \frac{1}{z_a^3 \, \sqrt{2}} \, \left[ z^4 - z_a^4
  \right].
\end{align}
Unfortunately the conditions in Eqs.~(\ref{free}) are insufficient to
fix $z_a$ uniquely.

However, $z_a$ in \eq{Psia} can be fixed if we choose to study a
closely relevant situation. Let us consider the trapped surface
formation in the collision of two sourced shock waves
\begin{subequations}
\begin{align}
  ds^2 \, = \, \frac{L^2}{z^2} \, \left\{ -2 \, dx^+ \, dx^- + \phi_1
    (z) \, \delta (x^-) \, d x^{- \, 2} + d x_\perp^2 + d z^2
  \right\} \\
  ds^2 \, = \, \frac{L^2}{z^2} \, \left\{ -2 \, dx^+ \, dx^- + \phi_2
    (z) \, \delta (x^+) \, d x^{+ \, 2} + d x_\perp^2 + d z^2
  \right\}
\end{align}
\end{subequations}
with the sources $J_{++}=\frac{E_1}{z_1 L} \, \delta(x^+) \,
\delta(z-z_1)$ and $J_{--}=\frac{E_2}{z_2 L} \, \delta(x^-) \,
\delta(z-z_2)$ corresponding to each of the shock waves. As discussed
in the previous sections, we keep $\frac{E_1 \, L^2}{z_1^4}=\frac{E_2
  \, L^2}{z_2^4}=\mu$ such that the nuclei on the boundary have the
same energy density.

The equations (\ref{free}) for the trapped surface now take the
following form:
\begin{subequations}\label{sourced}
\begin{align}
  &z^2\Psi_i''(z)-z\, \Psi_i'(z)-3 \, \Psi_i \, = \, -16 \, \pi \, G_5 \, E_i \, \delta(z-z_i)\\
  &\Psi_i(z_a) \, = \, \Psi_i(z_b) \, = \, 0\\
  &\frac{z_a^2}{L^2} \, \Psi_1'(z_a) \, \Psi_2'(z_a) \, = \,
  \frac{z_b^2}{L^2} \, \Psi_1'(z_b) \, \Psi_2'(z_b) \, = \, 8
\end{align}
\end{subequations}
where the boundary ${\cal C}$ is again $z_a<z<z_b$ and $i=1,2$.
\eq{sourced} is solved by
\begin{align}\label{Psi_s}
\Psi_i=\left\{\begin{array}{l@{\quad\quad}l}
C_i \, \left( \frac{z^3}{z_a^3}-\frac{z_a}{z} \right), & z<z_i\\~\\
D_i \, \left( \frac{z^3}{z_b^3}-\frac{z_b}{z} \right), & z>z_i
\end{array}
\right.
\end{align}
with the constants
\begin{align}\label{CD}
  \left\{\begin{array}{l}
      C_i=\frac{-4\pi G_5E_i}{z_i^4}\frac{\left( \frac{z_i^4}{z_b^4}-1 \right) \, z_b}{\frac{z_b^4-z_a^4}{z_a^3z_b^3}}\\~\\
      D_i=\frac{-4\pi G_5E_i}{z_i^4}\frac{\left( \frac{z_i^4}{z_a^4}-1
        \right) \, z_a}{\frac{z_b^4-z_a^4}{z_a^3z_b^3}}.
\end{array}
\right.
\end{align}
The third equation in (\ref{sourced}) gives the following simple relations:
\begin{align}\label{constr}
C_1 \, C_2 \, = \, D_1 \, D_2 \, = \, \frac{L^2}{2}.
\end{align}


\subsection{Shock Waves with Identical Sources}

It is instructive to first consider a collision of identical shock
waves in AdS$_5$. Putting $z_1 = z_2 = z_0$ and $E_1 = E_2 = E$ in
Eqs. (\ref{Psi_s}) and (\ref{CD}) above we obtain from \eq{constr}
\cite{Lin:2009pn}
\begin{align}\label{zab1}
\left\{ 
\begin{array}{c}
 z_a + z_b \, = \, \frac{4 \, \sqrt{2} \, \pi \, G_5}{L} \, E \\~~\\
 \frac{(z_a + z_b)^2 - 3 \, z_a \, z_b}{z_a^3 \, z_b^3} \, = \, \frac{1}{z_0^4}.
\end{array} \right.  
\end{align}
We want to take $z_0 \rightarrow \infty$ limit while keeping the
energy of the shock wave in the boundary theory fixed. That is we want
to hold 
\begin{align}
  \mu \, = \, \frac{E \, L^2}{z_0^4}
\end{align}
fixed. We rewrite \eq{zab1} in terms of $\mu$ as
\begin{align}\label{zab2}
\left\{ 
\begin{array}{c}
 z_a + z_b \, = \, \frac{2 \, \sqrt{2} \, \pi^2}{N_c^2} \, \mu \, z_0^4 \\~~\\
 \frac{(z_a + z_b)^2 - 3 \, z_a \, z_b}{z_a^3 \, z_b^3} \, = \, \frac{1}{z_0^4}
\end{array} \right.
\end{align}
where we have replaced $G_5 = \pi \, L^3 / 2 \, N_c^2$. Now, taking
$z_0 \rightarrow \infty$ keeping $\mu$ fixed we can easily infer the
asymptotics of $z_a$ and $z_b$. First one can consider the case that
in this limit $z_a$ and $z_b$ are of the same order, $z_a \sim z_b$.
In such case the first equation in (\ref{zab2}) gives $z_a \sim z_b
\sim z_0^4$, which can not satisfy the second equation in
(\ref{zab2}). As $z_a < z_b$ by definition, we are left to consider
the case when, in the $z_0 \rightarrow \infty$ limit one has $z_a \ll
z_b$. Then the first equation in (\ref{zab2}) yields
\begin{align}\label{zb}
  z_b \, \approx \, \frac{2 \, \sqrt{2} \, \pi^2}{N_c^2} \, \mu \,
  z_0^4
\end{align}
which, when plugged into the second equation in (\ref{zab2}) along
with the assumption that $z_a \ll z_b$ gives
\begin{align}\label{za}
  z_a \, \approx \, \frac{1}{\left( \frac{2 \, \sqrt{2} \,
        \pi^2}{N_c^2} \, \mu \right)^{\frac{1}{3}}} \, \equiv \,
  z_a^*.
\end{align}
The values of $z_a$ and $z_b$ given by Eqs. (\ref{za}) and (\ref{zb})
satisfy $z_a \ll z_b$ condition when $z_0$ is large, which confirms
that they give the correct asymptotics. In the strict $z_0 \rightarrow
\infty$ limit we see that $z_b \rightarrow \infty$, but $z_a$ remains
finite given by \eq{za}. Indeed Eqs. (\ref{za}) and (\ref{zb}) can
also be obtained by solving Eqs. (\ref{zab2}) explicitly and taking
the $z_0 \rightarrow \infty$ limit: the exact solution of \eq{zab2}
giving real $z_a$ and $z_b$ is
\begin{subequations}
\begin{align}
  z_a \, = \, \frac{{\tilde \mu} \, z_0^4}{2} - \frac{1}{4 \, \xi} \,
  \sqrt{ 2^{11/3} \, \xi^3 - 2^{13/3} \, z_0^4 \, \xi + 4 \, z_0^8 \,
    {\tilde \mu}^2 \, \xi^2} \label{zasol} \\
  z_b \, = \, \frac{{\tilde \mu} \, z_0^4}{2} + \frac{1}{4 \, \xi} \,
  \sqrt{ 2^{11/3} \, \xi^3 - 2^{13/3} \, z_0^4 \, \xi + 4 \, z_0^8 \,
    {\tilde \mu}^2 \, \xi^2} \label{zbsol}
\end{align}
\end{subequations}
where
\begin{align}
  {\tilde \mu} \, = \, \frac{2 \, \sqrt{2} \, \pi^2}{N_c^2} \, \mu
\end{align}
and
\begin{align}
  \xi \, = \, \left( z_0^6 \, \sqrt{4 + z_0^{12} \, {\tilde \mu}^4} -
    z_0^{12} \, {\tilde \mu}^2 \right)^{1/3}.
\end{align}
One can readily check that the $z_0 \rightarrow \infty$ asymptotics of
Eqs. (\ref{zasol}) and (\ref{zbsol}) is given by Eqs. (\ref{za}) and
(\ref{zb}).\footnote{Note that for $z_0 < (2/{\tilde \mu})^{1/3}$
  both $z_a$ and $z_b$ from Eqs. (\ref{zasol}) and (\ref{zbsol})
  become complex and trapped surface ceases to exist: however this
  small-$z_0$ limit is the exact opposite of the $z_0 \rightarrow
  \infty$ case we would like to consider here.}

Taking the $z_0 \rightarrow \infty$ limit in \eq{Psi_s} one can see
that the trapped surface is described by
\begin{align}
  \psi (z) \, = \, \frac{2 \, \pi^2}{N_c^2} \, \mu \, \left[ z^4 -
    z_a^{* \, 4} \right] \, = \, \frac{{\tilde \mu}}{\sqrt{2}} \,
  \left[ z^4 - {\tilde \mu}^{-4/3} \right],
\end{align}
which is exactly \eq{psia} with $z_a$ now fixed by \eq{za}. We see
that introducing bulk source as a regulator of the metric in the IR
and then taking $z_0 \rightarrow \infty$ limit allows one to fix $z_a$
and hence determines the trapped surface uniquely.

Now let us verify that the obtained value of $z_a$ in \eq{za} is
independent of the way we take the limit of sending the bulk sources
to infinite IR. Let us show that the same trapped surface arises in a
more general case when the two shock waves are different from each
other.


\subsection{Shock Waves with Sources at Different Bulk Locations}

We consider a collision of shock waves with sources at different
locations and with different $E_i$'s: now we have $z_1 \neq z_2$ and
$E_1 \neq E_2$ but with $\frac{E_1 \, L^2}{z_1^4}=\frac{E_2 \,
  L^2}{z_2^4}=\mu$. It proves useful to set
$\frac{z_1^4}{z_a^2z_b^2}=\lambda_1$,
$\frac{z_2^4}{z_a^2z_b^2}=\lambda_2$ and rewrite \eq{constr} as
\begin{align}\label{distinct}
  \left\{\begin{array}{l}
      \frac{z_a^2}{z_b^2}+\frac{z_b^2}{z_a^2}+1 = 
\frac{\lambda_1+\lambda_2+1}{\lambda_1\lambda_2} \\~\\
      (z_az_b)^3\frac{\frac{z_a}{z_b}+\frac{z_b}{z_a}}{\left( 
    \frac{z_a^2}{z_b^2}-\frac{z_b^2}{z_a^2} \right)^2} = \left( 
\frac{N_c^2}{2\pi^2\mu} \right)^2 \, \frac{1}{2(1-\lambda_1\lambda_2)} 
\end{array}
\right.
\end{align}
eliminating $z_1$ and $z_2$.  Finding solution for $z_a$ and $z_b$
seems to be a hard task. We instead first solve the first equation in
(\ref{distinct}) for $z_a/z_b$ and use the obtained ratio in the
second equation in (\ref{distinct}) to find $z_a \, z_b$. Using the
product $z_a \, z_b$ in $\frac{z_1^4}{z_a^2z_b^2}=\lambda_1$,
$\frac{z_2^4}{z_a^2z_b^2}=\lambda_2$ we can write $z_1$ and $z_2$ as
($i=1,2$)
\begin{align}\label{lam}
  z_i^4 \, = \, \lambda_i \, \left( \frac{N_c^2}{2 \, \pi^2 \, \mu}
  \right)^{4/3} \, \left[ \frac{\lambda_1 + \lambda_2 + 1 - 3 \,
      \lambda_1 \, \lambda_2}{2 \, (1-\lambda_1 \, \lambda_2)}
  \right]^{2/3} \, \frac{[(\lambda_1+1) \,
    (\lambda_2+1)]^{1/3}}{\lambda_1 \, \lambda_2}.
\end{align}
Again we have replaced $G_5 = \pi \, L^3 / 2 \, N_c^2$.  We are
interested in the limit $z_1,z_2\rightarrow\infty$ while keeping
$r_{12} = \frac{z_1}{z_2} = \text{finite}$ and $\mu$ is fixed. It is
not difficult to see that the limit can be achieved by taking
$\lambda_1,\lambda_2 \rightarrow 0$. In this limit \eq{lam} takes a
very simple form:
\begin{subequations}
\begin{align}
  z_1^4 \, & = \, \frac{1}{4 \, \lambda_2} \, \left(\frac{N_c^2}{\pi^2
      \, \mu}
  \right)^{4/3}\\
  z_2^4 \, & = \, \frac{1}{4 \, \lambda_1} \, \left( \frac{N_c^2}{\pi^2
      \, \mu} \right)^{4/3}.
\end{align}
\end{subequations}
As $z_b > z_a$, the first equation in (\ref{distinct}) gives in the
$\lambda_1,\lambda_2 \rightarrow 0$ limit that $z_b \gg z_a$. Solving
the second equation in (\ref{distinct}) for $z_b \gg z_a$ one obtains
$z_a$ asymptotics. Using the result in $\frac{z_1^4}{z_a^2 \,
  z_b^2}=\lambda_1$ along with the first equation in (\ref{distinct})
yields
\begin{subequations}
\begin{align}
\label{za2}&z_a \, \approx \, \left( \frac{N_c^2}{2 \, \sqrt{2} \, \pi^2 \, \mu} 
\right)^{1/3} \, \equiv \, z_a^*\\
\label{zb2}&z_b \, \approx \, (z_1z_2)^2 \, \left( \frac{N_c^2}{2\sqrt{2}\pi^2\mu} 
\right)^{-1}\rightarrow \infty .
\end{align}
\end{subequations}
These equations are completely analogous to Eqs. (\ref{za}) and
(\ref{zb}) above. Therefore the trapped surface is independent of the
way one send the bulk sources to the IR infinity: the sources do not
have to be at the same bulk location to obtain the same answer as we
had in the previous Subsection. To further test the independence of
taking the limit of sources going to infinite IR bulk, we have also
taken a similar limit for the point-like sources, first advocated in
\cite{Gubser:2008pc}: the trapped surface found in
\cite{Gubser:2008pc} again reduced to the trapped surface found in
this work above.

This completes our analysis of trapped surface in the collision of two
shock waves with sources infinitely deep in the bulk. We note unlike
source in finite depth\cite{Lin:2009pn}, no critical value for the
energy density is found in the limit. The formation of the trapped
surface is always guaranteed. The trapped surface is even independent
of the ratio $r_{12}=\frac{z_1}{z_2}$, i.e. the details of the limit!
It is important to stress that the trapped surface does not disappear
with the removal of the sources in the bulk, which can be viewed as IR
regulators. In all the examples of scattering of shock waves with bulk
sources the trapped surface always appears to be more or less centered
around the source in the $x_\perp, z$ space.  One was tempted to
conjecture therefore that the trapped surface is an inherent property
of non-zero bulk energy-momentum tensor. Our result proves otherwise,
giving an example of the source-free shock waves collision with a well
defined trapped surface.


\subsection{Limiting Trapped Surface}

To summarize our trapped surface analysis let us re-state that the
profiles of the trapped surface are given by
\begin{align}
  \Psi_i(z) \, = \, \frac{2 \, \pi^2 \, L \, \mu}{N_c^2} \, z_a^{* \,
    3} \, \left(\frac{z^3}{z_a^{* \, 3}}-\frac{z_a^*}{z} \right)
\end{align}
which is exactly \eq{Psia} with $z_a^*$ from \eq{za}. In the
transformed light cone coordinates (see \eq{coordtr}) the trapped
surface is then determined by
\begin{align}\label{fintr}
  x^+ \, = \,0, \ x^- \, = \, - \frac{\pi^2}{N_c^2} \, \mu \, \left[
    z^4 - z_a^{* \, 4} \right] \, = \, - \frac{{\tilde \mu}}{2 \,
    \sqrt{2}} \, \left[ z^4 - {\tilde \mu}^{-4/3} \right]
\end{align}
with an analogous expression for the other shock wave obtained by
interchanging $x^+ \leftrightarrow x^-$ in \eq{fintr}.
 
The trapped surface for a collision of source-free shock waves from
\eq{fintr} is illustrated in \fig{surf}. One can clearly see that the
trapped surface is present at all times before the collision and rises
from the deep IR toward finite values of $z$. Similar behavior was
observed for the trapped surface in the numerical model of heavy ion
collision involving gravitational perturbations in the 4-dimensional
world in \cite{Chesler:2008hg,Chesler:2009cy}. A horizon rising from the deep IR was
also deduced in \cite{Beuf:2009cx} for a model of heavy ion collision
involving a rapidity-independent matter distribution after the
collision.

\FIGURE{\includegraphics[width=12cm]{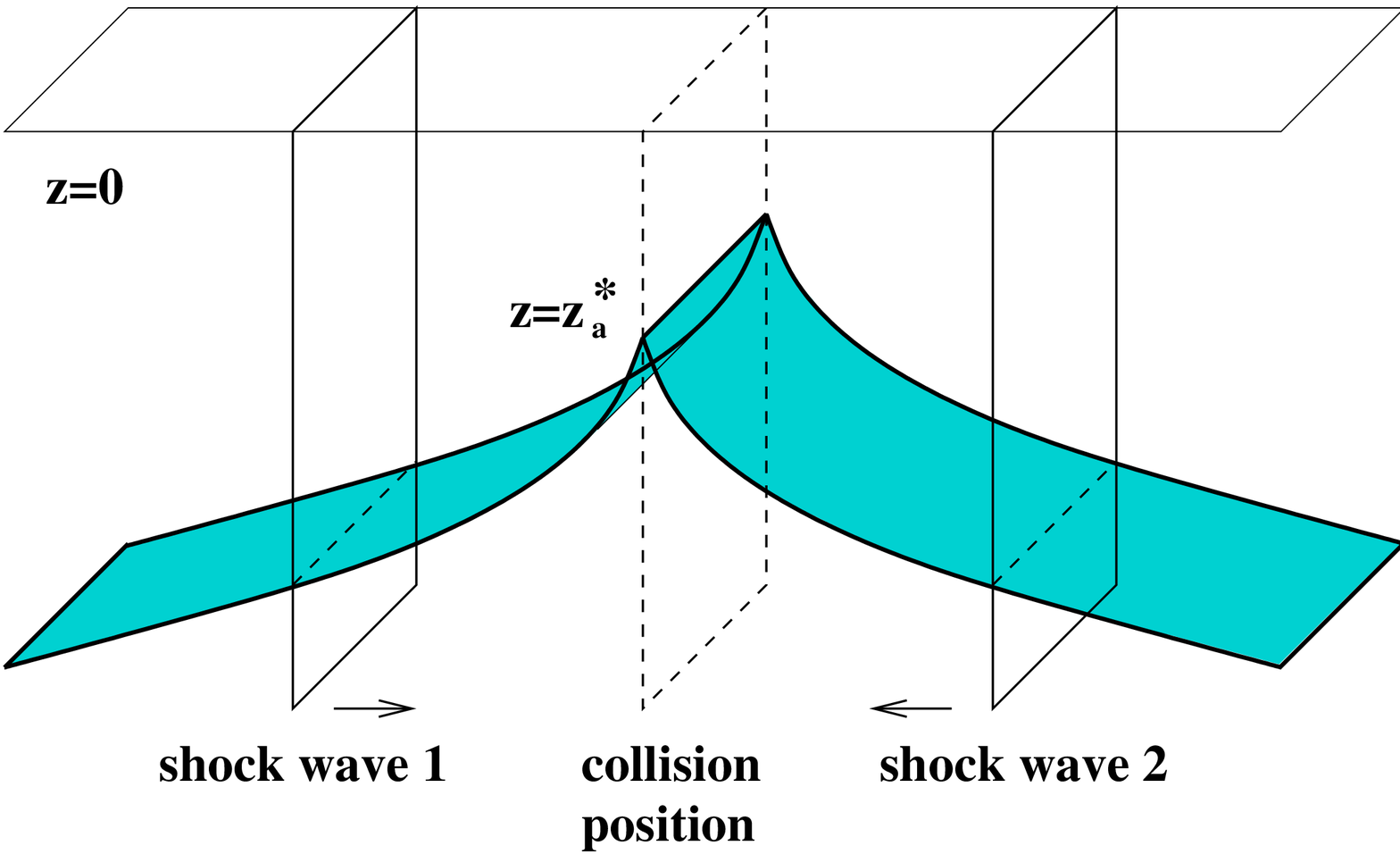}
\caption{An illustration of the trapped surface in the collision of 
  two sourceless shock waves. Vertical axis is the bulk $z$-direction,
  the horizontal left-right axis can be thought of either as the
  collision axis or as the time direction.  The trapped surface is
  shaded.}
\label{surf}
}

It is interesting to point out that the obtained shape of the trapped
surface appears to imply that the black hole produced in the collision
would have a singularity at $z=\infty$ with the horizon independent of
the transverse coordinates $x_\perp$. This is indeed very similar to
the black hole dual to Bjorken hydrodynamics constructed in
\cite{Janik:2005zt}. The main difference is that in our case the
metric (and the energy-momentum tensor in the gauge theory) are
rapidity-dependent, as follows from explicit calculations of the
metric produced in shock wave collisions
\cite{Grumiller:2008va,Albacete:2008vs,Albacete:2009ji}.

Our estimate for the produced entropy per unit transverse area
$A_\perp$ for a collision of two shock waves with the sources at $z_0$
is \cite{Lin:2009pn}
\begin{align}\label{ent}
  \frac{S}{A_\perp} \, = \, \frac{N_c^2}{2 \, \pi} \, \left[
    \frac{1}{z_a^2} - \frac{1}{z_b^2} \right].
\end{align}
Using Eqs. (\ref{za}) and (\ref{zb}) we obtain for $z_0 \rightarrow
\infty$
\begin{align}
  \frac{S}{A_\perp} \, = \, \left[ \pi \, N_c^2 \, \mu^2
  \right]^{1/3}.
\end{align}
As $\mu^2 \sim s$ with $s$ the center of mass energy of the collision,
we get 
\begin{align}
  \frac{S}{A_\perp} \, \propto \, s^{1/3}
\end{align}
in agreement with the result obtained in \cite{Gubser:2008pc}. 
\FIGURE{\includegraphics[width=11cm]{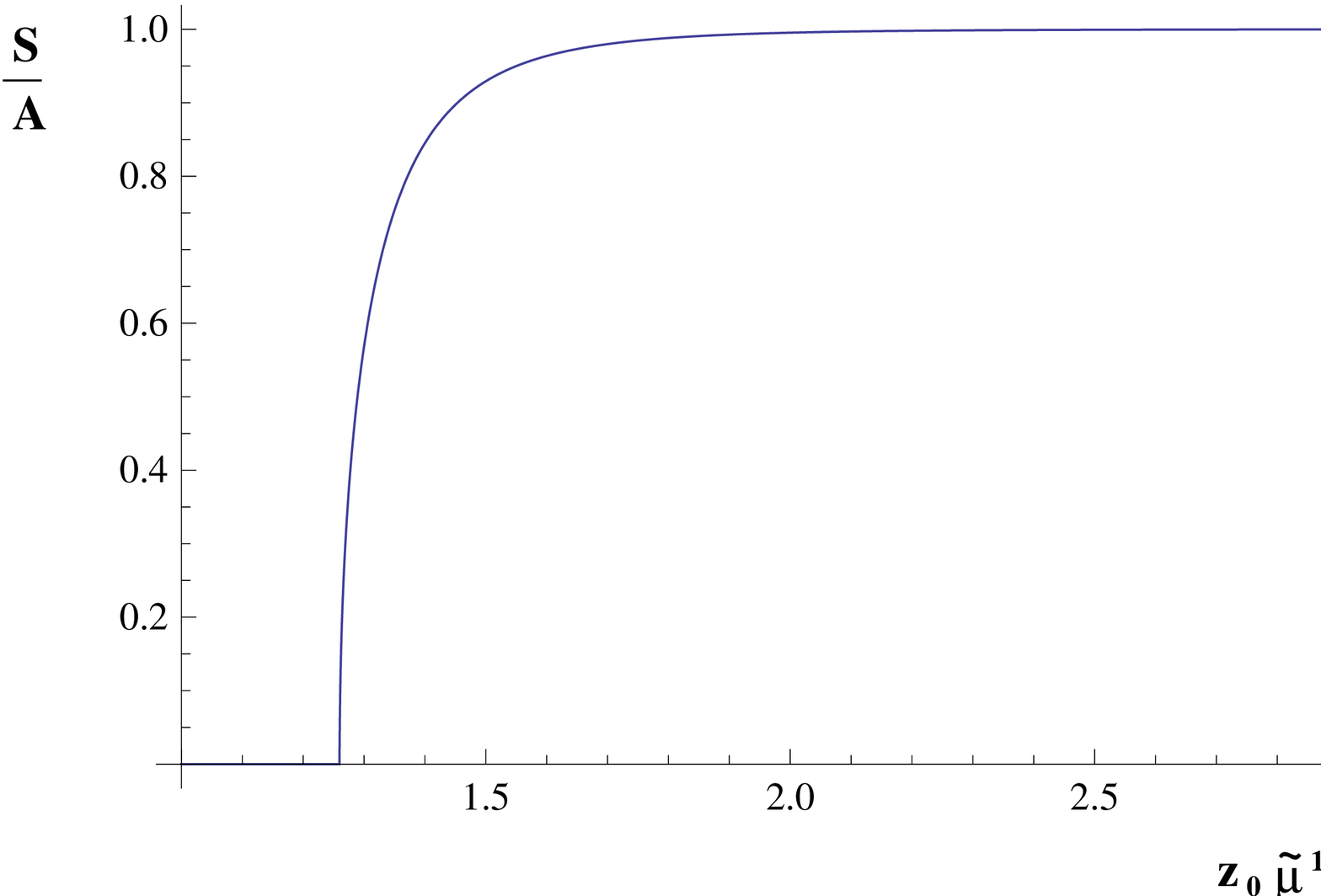}
\caption{The (lower bound on the) entropy density produced in the 
  collision of two identical shock waves with sources as a function of
  the source position $z_0$ in units of ${\tilde \mu}^{-1/3}$. The
  entropy density is in arbitrary units. }
\label{entropy}
}

The entropy from \eq{ent} is plotted in \fig{entropy} in arbitrary
units as a function of the bulk source location $z_0$ (for a collision
of two identical shock waves). \fig{entropy} demonstrated that
produced entropy becomes practically independent of the bulk source
position rather fast, approaching its asymptotic value well before
$z_0 \, {\tilde \mu}^{1/3}$ becomes large. 

As we noted above, for $z_0 \, {\tilde \mu}^{1/3} < 2^{1/3}$ both
$z_a$ and $z_b$ given by Eqs. (\ref{zasol}) and (\ref{zbsol}) become
complex and the trapped surface ceases to exist (see also
\cite{AlvarezGaume:2008fx} for a similar result). This likely implies
that no black hole is formed in collisions of such shock waves. To
understand this result from the boundary gauge theory perspective one
has to have a rigorous interpretation of what shock wave sources in
the bulk are dual to in the gauge theory. Such interpretation is
missing at the moment, which inspired our present investigation of
collisions of the sourceless shock waves. We may speculate though:
following \cite{Lin:2009pn} we may assume that the inverse position of
the source in the bulk $1/z_0$ provides an IR cutoff on the transverse
momenta of the partons in the shock waves' wave functions in the
boundary theory. Reducing $z_0$ would increase the cutoff $1/z_0$ thus
decreasing the number of partons: this is likely to lower the number
of degrees of freedom produced in the collision, leading to the
reduction of the entropy density with decreasing $z_0$ in
\fig{entropy}. Still it is not entirely clear why the trapped surface
disappears completely at a finite small $z_0$ forcing the estimate for
produced entropy to go to zero. Indeed our delta-function shock waves
are described by a single dimensionful parameter ${\tilde \mu}$ (or
$\mu$): the largest momentum scale in the problem is therefore
${\tilde \mu}^{1/3}$.  If $1/z_0$ is the IR cutoff, then clearly it
can not exceed the largest momentum scale: hence $1/z_0 \, \lsim \,
{\tilde \mu}^{1/3}$.  This, however, can not explain why the trapped
surface vanishes entirely at $z_0 = 2^{1/3} \, {\tilde \mu}^{-1/3}$.
Besides nothing pathological seems to happen in the perturbative
solution presented in Sect. \ref{pertsol} for small finite $z_0$. In
this work we are interested in the large-$z_0$ asymptotics:
\fig{entropy} demonstrates that the produced entropy density does not
seem to change much between having sources at finite large $z_0 >
2^{1/3} \, {\tilde \mu}^{-1/3}$ and having no sources at all, which
seems to agree with the IR cutoff interpretation of the sources and,
more importantly, shows that the entropy is ``well-behaved'' in the
$z_0 \rightarrow \infty$ limit we are taking. We leave the detailed
study of the small-$z_0$ regime for future work.

\section{Thermalization Time Estimate and Conclusions}
\label{conc}

The result fixing $z_a^*$ in \eq{za} could be predicted if one
realizes that in the limit of delta-function shock waves the problem
has only one dimensionful parameter ${\tilde \mu}$ which has
dimensions of mass cubed. If a non-vanishing trapped surface is
created in such collisions it has to be proportional to the only
distance scale in the problem: $1/{\tilde \mu}^{1/3}$. Stretching this
analogy further one should expect that the proper time of
thermalization (the time of black hole formation) is
\begin{align}\label{therm}
  \tau_{therm} \, \sim \, \frac{1}{{\tilde \mu}^{1/3}},
\end{align}
as was originally suggested in \cite{Grumiller:2008va}. 

An interesting question is the relation between this thermalization
time and the time it takes for the shock waves to stop. It was argued
in \cite{Albacete:2008vs,Albacete:2009ji} that colliding shock waves
come to a complete stop shortly after the collision. One can argue
that $\mu \sim p^+ \, \Lambda^2 \, A^{1/3}$ \cite{Albacete:2008vs},
where $p^+$ is the large longitudinal momentum of a ``nucleon'' in the
shock wave, $\Lambda$ is the typical transverse momentum scale in the
shock, and $A$ is the atomic number of the nucleus we model by the
shock wave. The characteristic light-cone stopping time for a shock
wave moving in the light-cone ``plus'' direction is given by
\cite{Albacete:2008vs,Albacete:2009ji}
\begin{align}\label{stop}
  x^+_{stop} \, \sim \, \frac{1}{\Lambda \, A^{1/3}}.
\end{align}
This is of course parametrically much longer than
\begin{align}\label{ttherm}
  \tau_{therm} \, \sim \, \frac{1}{{\tilde \mu}^{1/3}} \, \sim \,
  \frac{1}{(p^+ \, \Lambda^2 \, A^{1/3})^{1/3}}.
\end{align}
Hence, if one assumes that thermalization happens at mid-rapidity
first, then, as near mid-rapidity $t \approx \tau$, the time of
thermalization is $t_{therm} \approx \tau_{therm} \ll t_{stop} =
x^+_{stop}/\sqrt{2}$. It is therefore likely that thermalization
happens at times which are parametrically earlier than the stopping
time. If our guess of thermalization time is correct, this would imply
that the shock waves still move along their light cones when
thermalization happens, justifying an assumption commonly used in
hydrodynamic simulations of heavy ion collisions. Note also that the
thermalization time in \eq{ttherm} is very short, and decreases with
center-of-mass energy of the collision. (In fact, as was noticed in
\cite{Grumiller:2008va} this thermalization time is too short: if one
plugs in $p^+ =100$~GeV, $\Lambda = 0.2$~GeV and $A=196$ into the
parametric estimate (\ref{ttherm}) one would obtain $\tau_{therm}
\approx 0.07$~fm/c for RHIC, which is far too short for agreement with
hydrodynamic simulations \cite{Huovinen:2001cy,Teaney:2001av}. Indeed
the thermalization time estimate of \eq{therm} is too crude for
$0.07$~fm/c to be taken literally, and a numerical coefficient in
front of the estimate (\ref{therm}), if it results from a more exact
calculation and from a more realistic model of colliding nuclei, may
significantly change this number.)

It is important to stress the difference between the mathematical
limit of delta-function shock waves ($a \rightarrow 0$ with $a$ being
the $x^-$-width of the smeared non-delta-function shock wave
\cite{Albacete:2008vs,Albacete:2009ji} moving in the $x^+$ direction
or vice versa) and the physical high energy limit of $p^+ \gg \Lambda$
for nuclei.  While in the former limit ${\tilde \mu}$ is the only
non-vanishing dimensionful parameter in the problem, the latter limit
has another non-vanishing dimensionful scale ${\tilde \mu} \, a$,
which in fact gives the stopping time (\ref{stop}), $x^+_{stop} \,
\sim \, 1/\sqrt{{\tilde \mu} \, a}$
\cite{Albacete:2008vs,Albacete:2009ji}.  (As one can easily see $a
\sim A^{1/3}/p^+$ in the center-of-mass frame, such that ${\tilde \mu}
\, a \sim \Lambda^2 \, A^{2/3}$ is independent of $p^+$
\cite{Albacete:2008vs,Albacete:2009ji}.) With the presence of two
momentum scales in the problem the validity of the thermalization time
estimate of \cite{Grumiller:2008va} shown here in \eq{therm} becomes
less apparent. Our trapped surface analysis resulting in \eq{za}
appears to indicate that it is the momentum scale which depend only on
${\tilde \mu}$ and not on $a$ that matters for thermalization, thus
providing new evidence to support the estimate in \eq{therm}. In other
words we show that if one neglects the smaller second momentum scale
${\tilde \mu} \, a$ and approximates the shock wave profiles by
delta-functions, thermalization is achieved in the collisions at the
time given in \eq{therm}. If one treats the problem more carefully and
includes the scale ${\tilde \mu} \, a$ by considering shock waves of
finite longitudinal spread \cite{Albacete:2008vs,Albacete:2009ji},
\eq{therm} is likely to get corrections with the relative suppression
factor being some positive power of ${\tilde \mu} \, a / {\tilde
  \mu}^{2/3}$, which is very small for high energy collisions, thus
leaving the estimate in \eq{therm} practically unchanged.

One may argue that the strongly-coupled dynamics of the ${\cal N}=4$
SYM medium produced in shock wave collisions may be similar to that of
strongly-coupled QCD medium. Then our conclusion of rapid
thermalization may be applicable to soft (non-perturbative, $k_T \sim
\Lambda_{QCD}$) modes in heavy ion collisions, which would thermalize
very quickly. Harder (perturbative) modes may then thermalize through
interactions with the soft non-perturbative thermal bath, though more
work is needed to justify such thermalization scenario and to modify
the thermalization time estimate (\ref{ttherm}) to take into account
perturbative dynamics.

Another interesting question would concern understanding the relation
between the rather quick thermalization in heavy ion collisions for
the theory at strong coupling argued here and the impossibility of
thermalization at weak coupling suggested in \cite{Kovchegov:2005ss}
by one of the authors.\footnote{Note that perturbative thermalization
  scenarios have been advocated in
  \cite{Baier:2000sb,Mrowczynski:1988dz,Arnold:2003rq}.} While further
research is needed to clarify this problem, the solution may have
already been suggested in \cite{Cornalba:2009ax,Aharony:2005bm}, where
the authors argue that it is possible that there is a critical value
$\lambda_c$ of 't Hooft coupling $\lambda$. For $\lambda > \lambda_c$
black hole formation is likely in high energy collisions. At the same
time, for $\lambda < \lambda_c$ the black hole is {\sl not} formed in
high energy collisions \cite{Cornalba:2009ax,Aharony:2005bm},
corresponding to no thermalization in the boundary theory. Indeed in
the case of real-life heavy ion collisions, due to the running of the
strong coupling constant, the coupling assumes a wide range of values
in a single collision. The coupling is always large for soft
transverse modes, making thermalization due to large coupling effects
likely in light of our above results.

To conclude let us point out once more that we have obtained a trapped
surface for a collision of two sourceless shock waves in AdS$_5$. The
shape of the trapped surface is given by \eq{fintr} and is illustrated
in \fig{surf}. Existence of this trapped surface proves that a black
hole is created in the bulk for a collision of two sourceless shock
waves, corresponding to creation of thermalized medium (quark-gluon
plasma) in the boundary gauge theory.



\acknowledgments

The authors would like to thank Edward Shuryak and Anastasios Taliotis for discussions.

The work of Yu.K. is sponsored in part by the U.S. Department of
Energy under Grant No. DE-FG02-05ER41377. The work of S.L. is partially
supported by the US-DOE grants DE-FG02-88ER40388 and
DE-FG03-97ER4014.



\providecommand{\href}[2]{#2}\begingroup\raggedright\endgroup


\end{document}